\documentclass[aps,prd,reprint,superscriptaddress,showkeys,amsmath,amssymb,amsfonts]{revtex4-2}
\usepackage[T1]{fontenc}
\usepackage[utf8]{inputenc}\DeclareUnicodeCharacter{2212}{-}
\usepackage[spanish,USenglish]{babel}
\usepackage[colorlinks,citecolor=blue,urlcolor=blue,linkcolor=blue]{hyperref}
\usepackage{braket}
\usepackage{mathtools}
\usepackage{enumerate}
\usepackage{bm}
\usepackage{dcolumn}
\usepackage{pgf}

\DeclarePairedDelimiter{\abs}{\lvert}{\rvert}

\begin{document}

\title{Diabatic description of bottomoniumlike mesons}
\author{R. Bruschini}
\email{roberto.bruschini@ific.uv.es}
\affiliation{\foreignlanguage{spanish}{Unidad Teórica, Instituto de Física Corpuscular (Universidad de Valencia--CSIC), E-46980 Paterna (Valencia)}, Spain}
\author{P. González}
\email{pedro.gonzalez@uv.es}
\affiliation{\foreignlanguage{spanish}{Unidad Teórica, Instituto de Física Corpuscular (Universidad de Valencia--CSIC), E-46980 Paterna (Valencia)}, Spain}
\affiliation{\foreignlanguage{spanish}{Departamento de Física Teórica, Universidad de Valencia, E-46100 Burjassot (Valencia)}, Spain}


\begin{abstract}
We apply the diabatic approach, specially suited for a QCD based study of
conventional (quark-antiquark) and unconventional (quark-antiquark + meson-meson)  meson states,
to  the description of hidden-bottom mesons. A spectral analysis of the $I=0$,
$J^{++}$ and $1^{--}$ resonances with masses up to about $10.8$ GeV is carried out.
Masses and widths of all the experimentally known resonances, including
conventional and unconventional states, can be well reproduced. In particular, we predict a significant $B\bar{B}^{\ast}$
component in $\Upsilon(10580)$.
We also predict the existence of a not yet discovered unconventional  $1^{++}$ narrow state, with a significant
$B_{s}\bar{B}_{s}^{\ast}$ content
making it to decay into $\Upsilon(1S)\phi$, whose experimental discovery
would provide definite support to our theoretical analysis.
\end{abstract}

\maketitle

\section{Introduction\label{SI}}

The unified description of conventional and unconventional heavy-quark mesons
from QCD, the strong interaction theory, is a current theoretical challenge in
hadron physics. Due to the current impossibility of solving QCD in the
nonperturbative regime, effective field theories directly connected to QCD,
involving quark and gluons or hadrons as degrees of freedom, have been
developed for the study of the heavy-quark meson structure, see for instance
\cite{Bra20} and references therein. On the other hand, QCD calculations of
heavy-quark mesons in the lattice have been performed. These comprise quenched
analyses involving $Q\bar{Q}$ $(Q:\text{ heavy quark, }b\text{ or
}c)$ with gluons as the light field \cite{Jug99,Bal01}, and unquenched
studies with $Q\bar{Q}$ and meson-meson components incorporating also
light sea quarks in the light field \cite{Bal05,Bal11,Bul19}. A nice feature
of the lattice, concerning phenomenology, is that it provides a straightforward
way to compute complete heavy-quark meson potentials from QCD: the static
light field energies evaluated in lattice are related to static potentials.
More concretely, following a Born-Oppenheimer approximation quenched static
energies can be directly identified with potentials in a Schr\"{o}dinger
equation for $Q\bar{Q}$, see for instance \cite{Jug99,Braa14}. This
allows for a QCD-based description of conventional quarkonium $(
b\bar{b}\text{ or }c\bar{c})$ in terms of a potential whose
spin-independent part corresponds to a Cornell (funnel) form. As for
unquenched static energies, calculated for $Q\bar{Q}$ in the presence of
meson-meson configurations, the Born-Oppenheimer approximation, which is a
single channel one, is not valid anymore. Instead, a diabatic approach
\cite{Bru20} permits their connection with the potential matrix in a
multichannel Schr\"{o}dinger equation for the $Q\bar{Q}$ and meson-meson components.

Strictly speaking the static potential is only exact in the limit of infinite
heavy-quark mass. For bottomonium $(b\bar{b})$ with a
quark mass, $m_{b}$, much larger than the QCD scale, $\Lambda_{QCD}$, the
static limit represents a rather good approximation. For charmonium ($
c\bar{c}$), with a much lower quark mass, $m_\textup{c}$, nonstatic
contributions could be significant. Despite this drawback, in the last two
decades, much more attention has been paid to the theoretical description of
the excited spectrum of charmonium, the reason being the discovery, starting
at 2003 with the $\chi_{c1}(3872)$, of charmoniumlike mesons
whose properties (masses and widths) cannot be properly described from a
conventional $c\bar{c}$ structure. The role played by explicit or
implicit open charm meson-meson components in the description of these
unconventional states has been recognized, and alternative
models\ (meson-meson molecules, tetraquarks, hadrocharmonium) have been
formulated, some reviews are \cite{Guo18,Leb17,Esp17,Chen16,Vol07}.

Quite recently, a (nonperturbative) diabatic description of the $I=0$,
$J^{++}$ and $1^{--}$ hidden-charm mesons with masses up to about $4$ GeV, in
terms of $c\bar{c}$ and meson-meson components, has been undertaken
\cite{Bru20,Bru21}. A major difference with respect to other nonperturbative
studies involving the same degrees of freedom, see for example
\cite{Eic04,Eic06}, is the incorporation of a lattice-based form of the mixing
potential instead of an ansatz with no clear connection to QCD. Despite the
dearth of lattice data, and the technical approximations followed for tackling
the diabatic equations, the results obtained (masses and widths) are
encouraging. This supports the diabatic approach in QCD as an appropriate
framework for a unified and complete nonperturbative description of
conventional and unconventional heavy-quark mesons.

For hidden-bottom mesons there have been in the past many speculations about
possible bottomoniumlike partners of the unconventional charmoniumlike states,
see for instance \cite{Ols15} and references therein. The partner hypothesis
is based on the consideration that hidden-charm and hidden-bottom mesons can be
described from the same flavor independent static potential (up to a
constant). This is clearly acceptable for (conventional) charmonium,
$c\bar{c}$, and bottomonium, $b\bar{b}$, with masses lying below the
first open-flavor meson-meson threshold, which are quite successfully
described from a quark-antiquark Cornell potential. However, for
unconventional states involving $Q\bar{Q}$ and open-flavor meson-meson
components as well, the partner hypothesis is questionable, for it is doubtful that the off-diagonal
terms in the static potential matrix, giving account of the
$Q\bar{Q}$ and meson-meson mixings, be flavor independent.
Experimentally, the situation is not well established due to the
current dearth of data (masses and widths) for $I=0$, $J^{++}$ hidden-bottom
mesons above the first $(0^{++})$ open-bottom meson-meson
threshold, and the absence of data for $I=0$, $1^{--}$ resonances with
masses above the first $1^{--}$ $S$-wave meson-meson threshold. From the
theoretical point of view, the diabatic approach, generating the static
potential matrix from lattice QCD data, can be an ideal tool to definitely settle
this issue. Indeed, lattice data for the energy of static $b$ and
$\bar{b}$ sources, when the $b\bar{b}$ configuration mixes with one
or two open-flavor meson-meson ones, are available. From them a direct
parametrization of the diabatic potential matrix is possible, and a QCD based
prediction of the unknown excited spectrum is feasible. Actually, the
hidden-bottom meson spectrum has been partially explored recently in a
simplified diabatic treatment of $I=0$, $0^{-+}$ and $1^{--}$ resonances,
involving only one $b\bar{b}$ channel and at most two distinct
meson-meson thresholds masses\cite{Bic20,Bic21}.

In this article we center on the diabatic description of hidden-bottom mesons.
The main differences with respect to \cite{Bic20,Bic21} are i) the 
consideration of all possible $b\bar{b}$ channels and  all meson-meson
threshold masses contributing, ii) the mixing potential which in our case does not
contain any short range (light quark meson exchange) contribution, in line with
the use of constant meson-meson potentials, and iii) the use of a bound-state based
approximation instead of a $\mathrm{S}$-matrix approach to the spectral solutions.
We restrict our study to $I=0$,
$J^{++}$ and $1^{--}$ resonances with masses up to about $10.8$ GeV, two
hundreed of MeV below the first $1^{--}$ $S$-wave meson-meson thresholds.
Thus, as all the lower thresholds are known and have very small widths we
avoid the uncertainty deriving from the partial knowledge of a threshold and
the complexity due to possible threshold width effects. For the the sake of
technical simplicity in the evaluation of observables, we follow a two step description of resonances: first we
approximate them by stable bound states incorporating closed meson-meson
channels, and second we calculate mass corrections and widths from open
meson-meson channels. We show that a fairly good description of the currently
known $J^{++}$ and $1^{--}$ experimental resonances in the realm of energy
under study comes out. We predict that all these resonances except
$\Upsilon(10580)$ have a very predominant $b\bar{b}$
component. For $\Upsilon(10580)$ the reduced, albeit dominant,
$b\bar{b}$ probability allows to give accurate account of leptonic width
data. As for the not yet discovered resonances we predict that only for the
third excited $1^{++}$ state there is a significant meson-meson component.
Although not dominant, this component points out to $\Upsilon(
1S)\phi$ as a favored decay channel what could be relevant for its
experimental discovery. Altogether these results indicate that a
partner picture of hudden-charm and hidden-bottom mesons should be discarded once meson-meson
contributions start to play some role.

These contents are organized as follows. In Sec.~\ref{SII} a brief review of the
diabatic formalism particularized for hidden-bottom mesons is presented, and
the diabatic potential matrix is built from lattice data. As an improvement
over the previous development for hidden-charm mesons a distinctive treatment
of hidden-strange thresholds is incorporated. In Sec.~\ref{SIII} the
nonperturbative description of hidden-bottom mesons is done in two steps:
first, a bound state approximation incorporating closed meson-meson thresholds
is followed, and second, mass shifts and widths from open meson-meson
thresholds are calculated. Finally, in Sec.~\ref{SIV} our main results and
conclusions are summarized.

\section{Diabatic formalism for hidden-bottom mesons\label{SII}}

The diabatic approach in QCD has been developed in \cite{Bru20}. Hidden-bottom
meson states with quantum numbers $J^{PC}$, made of $b\bar{b}$ and open-bottom
meson-meson $M_{1}^{(i)}\bar{M}_{2}^{(i)}$ components, with
$M_{1}$ $(\bar{M}_{2})$ containing $q\bar{b}$
$(\bar{q}b)$ where $q$ stands for a light quark,
$q=u,d,s$, are solutions of the multichannel Schr\"{o}dinger equation%
\begin{equation}
(\mathrm{K}+\mathrm{V}(\bm{r}))\Psi(\bm{r}%
)=E\Psi(\bm{r})
\end{equation}
where $\Psi(\bm{r})$ is a column vector%
\begin{equation}
\Psi(\bm{r})=%
\begin{pmatrix}
\psi_{b\bar{b}}(\bm{r})\\
\psi^{(1)}(\bm{r})\\
\vdots\\
\psi^{(N)}(\bm{r})\\
\end{pmatrix}
\end{equation}
with $\psi_{b\bar{b}}(\bm{r})$ standing for the
$b\bar{b}$ component, and $\psi^{(i)}(\bm{r})$, $i=1,2\dots$
for the $M_{1}^{(i)}\bar{M}_{2}^{(i)}$ component.

$\mathrm{K}$ is the kinetic energy matrix%

\begin{equation}
\mathrm{K}=%
\begin{pmatrix}
-\frac{1}{2\mu_{b\bar{b}}}\nabla^{2} &  &  & \\
& -\frac{1}{2\mu^{(1)}}\nabla^{2} &  & \\
&  & \ddots & \\
&  &  & -\frac{1}{2\mu^{(N)}}\nabla^{2}%
\end{pmatrix}
\label{Kinetic}%
\end{equation}
where $\mu_{b\bar{b}}$ is the reduced $b\bar{b}$ mass, $\mu^{(i)}$
is the reduced $M_{1}^{(i)}\bar{M}_{2}^{(i)}$ mass, and matrix elements
equal to zero are not displayed.

$\mathrm{V}(\bm{r})$ is the diabatic potential matrix. Up to spin
dependent terms that we shall not consider it can be formally written as%
\begin{equation}%
\begin{pmatrix}
V_{\text{C}}(r) & V_{\textup{mix}}^{(1)}(r) & \hdots & V_{\textup{mix}}%
^{(N)}(r)\\
V_{\textup{mix}}^{(1)}(r) & T^{(1)} &  & \\
\vdots &  & \ddots & \\
V_{\textup{mix}}^{(N)}(r) &  &  & T^{(N)}%
\end{pmatrix}
\end{equation}
where the diagonal elements $V_{\text{C}}(r)$ and $T^{(i)}$
correspond to the $b$-$\bar{b}$ and $M_{1}^{(i)}$-$\bar{M}_{2}^{(i)}$
potentials respectively, and $V_{\textup{mix}}^{(i)}(r)$ to the $M_{1}^{(i)}\bar{M}_{2}^{(i)}$-$b\bar{b}$
interaction potential.

More precisely, we express the $b\bar{b}$ component as%
\begin{equation}
\psi_{b\bar{b}}(\bm{r})=\sum_{t}%
R_{t}^{(0)}(r)
\mathcal{Y}_{l_{t}^{(0)},s_{t}^{(0)}}^{J,m_{J}%
}(\bm{\hat{r}})
\end{equation}
where the sum over $t$ goes from $1$ to the number of pairs $(
l_{b\bar{b}}\equiv l^{(0)},s_{b\bar{b}}\equiv
s^{(0)})$ coupling to $J^{PC}$, $R_{t}^{(0)}(r)$ stands for a radial wavefunction and
\begin{equation}
\mathcal{Y}_{l,s}^{J,m_{J}}(\widehat{r})\equiv\sum_{m_{l}%
,m_{s}} C_{l, s, J}^{m_{l}, m_{s}, m_{J}} Y_{l}^{m_{l}}(\bm{\hat{r}})
\xi_{s}^{m_{s}}
\end{equation}
for an angular-spin wavefunction ($C$ is a Clebsch-Gordan coefficient,
$Y_{l}^{m_{l}}$ a spherical harmonic and $\xi_{s}^{m_{s}}$ a spin vector), and the
$M_{1}^{(i)}\bar{M}_{2}^{(i)}$ component as
\begin{equation}
\psi^{(i)}(\bm{r})=\sum_{k}%
R_{k}^{(i)}(r)
\mathcal{Y}_{l_{k}^{(i)},s_{k}^{(i)}}^{J,m_{J}} (\bm{\hat{r}})
\end{equation}
where the sum over $k$ goes from $1$ to the number of pairs $(
l^{(i)}\equiv l_{M_{1}^{(i)}\bar{M}_{2}^{(i)}},s^{(
i)}\equiv s_{M_{1}^{(i)}\bar{M}_{2}^{(i)}})$ coupling to
$J^{PC}$.

(Let us note that we have changed the notation for the radial wavefunction
with respect to our previous papers \cite{Bru20,Bru21}. Here we use the
standard $R$ and reserve $u$ for the reduced radial wavefunction, see next.)

Then, one has
\begin{subequations}
\begin{align}
\int \mathrm{d}\Omega \, \psi_{b\bar{b}}^{\ast}(\bm{r})
\mathrm{V}(\bm{r})\psi_{b\bar{b}}(\bm
{r}) =& \sum_{t}R_{t}^{(0)\ast}(r)V_{\text{C}}(r)R_{t}^{(0)}(r) \\
\int \mathrm{d}\Omega \, \psi^{(i)\ast}(\bm{r})
\mathrm{V}(\bm{r})\psi_{b\bar{b}}(\bm
{r}) =& \sum_{k, t} R_{k}^{(i)\ast}(r) V_{\textup{mix}}^{(i)}(r) R_{t}^{(0)}(r) \\
\int \mathrm{d}\Omega \,\psi^{(i^{\prime}) \ast}(\bm
{r})\mathrm{V}(\bm{r})\psi^{(i)}(
\bm{r}) =& \delta_{i i^{\prime}}\sum_{k}R_{k}^{(i)\ast}(r)T^{(i)}
R_{k}^{(i)}(r)
\end{align}
\end{subequations}
so that the multichannel Schr\"{o}dinger equation reduces to a coupled system
of radial equations for the sets of channels $\bigl\{  u_{t}^{(0)}(r)\equiv r R_{t}^{(0)}(r)\bigr\}$ and
$\bigl\{u_{k}^{(i)}(r)\equiv r R_{k}^{(i)}(r)\bigr\}$.

For example, if we considered for simplicity the case of the $b\bar{b}$
component with only one pair, $(l_{1}^{(0)}%
,s_{1}^{(0)})$, coupling to the given $J^{PC}$, and one
meson-meson component $M_{1}^{(1)}\bar{M}_{2}^{(1)}$ with only one pair,
$(l_{1}^{(1)},s_{1}^{(1)})$,
coupling to the given $J^{PC}$, the system would read
\begin{widetext}
\begin{equation}
\begin{pmatrix}
-\frac{1}{2\mu_{b\bar{b}}}(\partial_{r}^{2}-\frac{l_{1}^{(
0)}(l_{1}^{(0)}+1)}{r^{2}})
+V_{\text{C}}(r)-E & V_{\textup{mix}}^{(1)}(r)\\
V_{\textup{mix}}^{(1)}(r) & -\frac{1}{2\mu^{(1)}}(\partial_{r}%
^{2}-\frac{l_{1}^{(1)}(l_{1}^{(1)
}+1)}{r^{2}})+T^{(1)}-E
\end{pmatrix}
\begin{pmatrix}
u_{1}^{(0)}\\
u_{1}^{(1)}
\end{pmatrix}=0.
\end{equation}
\end{widetext}
The generalization to any number of possible $(l^{(0)
},s^{(0)})$ and $(l^{(i)
},s^{(i)})$, $i=1,2..$. pairs is straightforward by
considering each $u_{t}^{(0)}$
and each $u_{k}^{(i)}$ as a
component of the eigenfunction. Then, for normalizable solutions of the general system
of radial equations, the probability for the $b\bar{b}$ component can be
calculated as
\begin{equation}
\mathcal{P}(b\bar{b})=\sum_{t}\int \mathrm{d}r \abs{u_{t}^{(0)}(r)}^{2}%
\end{equation}
and for the $M_{1}^{(i)}\bar{M}_{2}^{(i)}$ component
\begin{equation}
\mathcal{P}(M_{1}^{(i)}\bar{M}_{2}^{(i)})=\sum_{k}\int
\mathrm{d}r\abs{u_{k}^{(i)}(r)}^{2}.
\end{equation}


Notice that although no direct interaction potential between different
meson-meson components is considered, what it is justified for isolated, well
separated meson-meson thresholds with no overlap at all, an indirect
interaction through their coupling to the $b\bar{b}$ channel is present.

\subsection{Diabatic potential matrix}

The explicit form of the matrix elements $V_{\text{C}}(r)$, $T^{(
i)}$, $V_{\textup{mix}}^{(i)}(r)$ can be derived from the light field
static energies calculated in lattice QCD \cite{Bru20}. As lattice results
depend on the chosen lattice spacing the philosophy underlying this derivation
is the use of parametrizations motivated from lattice results with parameters
to be fixed from phenomenology. Thus, the diagonal element $V_{\text{C}}(r)$
corresponding to the $b$-$\bar{b}$ potential is parametrized from
quenched lattice data on the static quark-antiquark energy \cite{Bal01} as the
Cornell potential
\begin{equation}
V_{\text{C}}(r)=\sigma r-\frac{\chi}{r}-\beta+ m_{b}+m_{\bar{b}}
\label{CPOT}
\end{equation}
with $\sigma$, $\chi$, $\beta$ and $m_{b}$ being the string tension, the
colour coulomb strength, a constant, and the bottom quark mass respectively. We
shall assume that all the flavor dependence in $V_{\text{C}}(r)$ comes from
the mass term. Therefore, we shall keep
for hidden-bottom mesons the same values for $\sigma$, $\chi$, and $\beta$ used
in \cite{Bru20} for hidden-charm mesons. In order to fix $m_{b}$ we have to
take into account that the potential is spin independent so that the
calculated masses should be compared with the experimental mass centroids
obtained from spin singlet and spin triplet data. So we choose to fit the $1P$
ground state mass centroid under the assumption that $1P_{J}$ experimental
resonances are pure bottomonium states, as will be confirmed later on
(alternatively we could have chosen to fit the $1S$ or $2S$ or $2P$
mass centroid without any significant change in the forthcoming analysis).
Thus, we have
\begin{subequations}\begin{align}
\sigma &  =925.6\text{~MeV/fm},\\
\chi &  =102.6\text{~MeV~fm},\\
\beta &  = 855\text{~MeV}.\\
m_{b} &  =5215\text{~MeV}.
\end{align}\end{subequations}
The $b\bar{b}$ spectrum from this Cornell potential for $J^{++}$ and
$1^{--}$ isoscalar states is shown in Table~\ref{tabcor}.

\begin{table}
\caption{Bottomonium spectrum from the Cornell potential. Each spectral state is characterized by $J^{PC}$ and $nL$ quantum numbers. For $J^{PC}=(0,1,2)^{++}$, it is intended that $F$-wave bottotmonium states appear only for $2^{++}$. Available experimental centroid masses from \cite{PDG20} are
listed for comparison.\label{tabcor}}
\begin{ruledtabular}
\begin{tabular}{ccdd}
$J^{PC}$			& $nL$	& {\text{Mass (MeV)}}	& {\text{Centroid (MeV)}}	\\
\hline
$(0, 1, 2)^{++}$	& $1P$ 	& 9900.7				& 9899.7				\\
				& $2P$ 	& 10254.4				& 10260.2				\\
				& $1F$ 	& 10341.5	\\
				& $3P$ 	& 10536.6	\\
				& $2F$ 	& 10601.0	\\
				& $4P$ 	& 10782.2	\\
$1^{--}$			& $1S$	& 9401.2				& 9444.9				\\
				& $2S$	& 9993.8				& 10017.2				\\
				& $1D$	& 10150.4	\\
				& $3S$	& 10338.6	\\
				& $2D$	& 10442.0	\\
				& $4S$	& 10615.0	\\
				& $3D$	& 10694.1	\\
				& $5S$	& 10856.4	\\
\end{tabular}
\end{ruledtabular}
\end{table}

Let us point out that in phenomenological applications of the Cornell
potential \cite{Eic94,Eic80} the chosen value of the bottom quark mass differs
slightly from ours. In these applications distinct values of $\beta$ are
considered for bottomonium and charmonium in order to fit approximately the
low-lying mass centroids.


Any of the other diagonal elements $T^{(i)}$ represents a $M_{1}^{(i)}$-$\bar{M}_{2}^{(i)}$ potential. Up to one pion exchange effects that
we do not consider this potential is given by the $i$th meson-meson
threshold
\begin{equation}
T^{(i)}=m_{M_{1}^{(i)}}+m_{\bar{M}_{2}^{(i)}}%
\end{equation}
with $m_{M_{1}^{(i)}}$ and $m_{\bar{M}_{2}^{(i)}}$ being the masses of
the corresponding mesons. The meson-meson thresholds, calculated from the
masses of bottom mesons in \cite{PDG20}, are listed in Table~\ref{tablist}.

\begin{table}
\caption{\label{tablist}Low-lying open-bottom meson-meson thresholds $M^{(i)}\bar{M}^{(i)}$.
Threshold masses $T^{(i)}$ from the bottom and bottom strange meson masses quoted in \cite{PDG20}. Crossing
radii of these thresholds with the Cornell potential, $r_\textup{c}^{(i)}$, are also tabulated.}
\begin{ruledtabular}
\begin{tabular}{cccc}
$i$	& $M_{1}^{(i)}\bar{M}_{2}^{(i)}$	& $T^{(i)}$ (MeV)	& $r_\textup{c}^{(i)}\text{~(fm)}$	\\
\hline
$1$	& $B\bar{B}$ 					& $10559$			& $1.16$						\\
$2$	& $B\bar{B}^{\ast}$ 				& $10604$			& $1.20$						\\
$3$	& $B^{\ast}\bar{B}^{\ast}$		& $10649$			& $1.25$						\\
$4$	& $B_{s}\bar{B}_{s}$				& $10733$			& $1.33$						\\
$5$	& $B_{s}\bar{B}_{s}^{\ast}$ 		& $10782$			& $1.38$						\\
$6$	& $B_{s}^{\ast}\bar{B}_{s}^{\ast}$	& $10830$			& $1.43$						\\
\end{tabular}
\end{ruledtabular}
\end{table}

It is worth remarking that the use of the experimental masses for the
thresholds introduces some implicit spin dependence in the description.


Let us note that each of the $B\bar{B}$, $B\bar{B}^{\ast}$ and $B^{\ast
}\bar{B}^{\ast}$ thresholds is composed of two (approximately) degenerate thresholds. For example
$B\bar{B}$ corresponds to $B^{+}B^{-}$ and $B^{0}\bar{B}^{0}$, with
an experimental threshold mass difference of $0.6$ MeV. In contrast the hidden
strange cases $B_{s}\bar{B}_{s}$, $B_{s}\bar{B}_{s}^{\ast}$ and
$B_{s}^{\ast}\bar{B}_{s}^{\ast}$ are single thresholds.


The off-diagonal elements, $V_{\textup{mix}}^{(i)}(r)$, correspond to
$b\bar{b}$-$M_{1}^{(i)}\bar{M}_{2}^{(i)}$ mixing potentials. From
unquenched lattice static energies, calculated for $b\bar{b}$ in the
presence of meson-meson configurations \cite{Bal05,Bul19}, the following
parametrization has been proposed \cite{Bru20}%

\begin{equation}
\abs{{V_{\textup{mix}}^{(i)}(r)}}=\frac{\Delta^{(i)}}{2}%
\exp\biggl\{-\frac{(V_{\text{C}}(r)-T^{(i)})^{2}}{2\sigma^{2}\rho^{2}}\biggr\}
\end{equation}
where $\rho$ is a radial scale for the mixing, that we shall take equal for
all thresholds, and $\Delta^{(i)}$ is a strength parameter
corresponding to the difference between the unquenched lattice static energies
resulting from the avoided crossing of $V_{\text{C}}(r)$ and $T^{(i)}$ at the
crossing radius $r_\textup{c}^{(i)}$ defined by%
\begin{equation}
V_{\text{C}}(r_\textup{c}^{(i)})=T^{(i)}.
\end{equation}
The values of the crossing radii have been tabulated in Table~\ref{tablist}.

For the sake of simplicity, in \cite{Bru20,Bru21} the same value for
$\Delta^{(i)}$ was used for degenerate and single thresholds.
Here we go a step further. As shown in the Appendix a doubly degenerate threshold can
be managed as an effective single threshold with a different value of $\Delta
$:
\begin{equation}
\Delta_{\text{degenerate}}=\sqrt{2}\Delta_{\text{single}}.
\end{equation}

\begin{figure}
\centering
\input{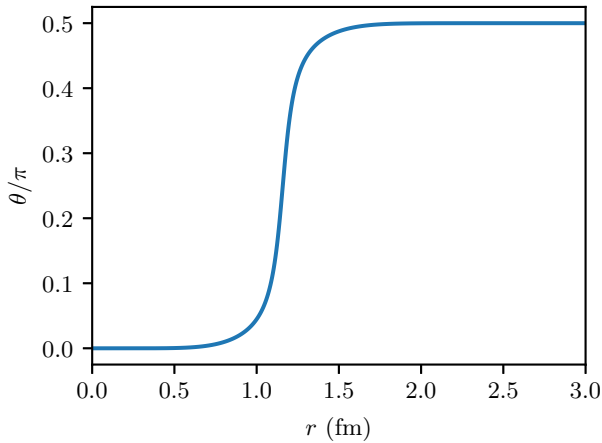}
\caption{Mixing angle between $b\bar{b}$ and $B \bar{B}$.\label{mixang}}
\end{figure}

\begin{figure}
\centering
\input{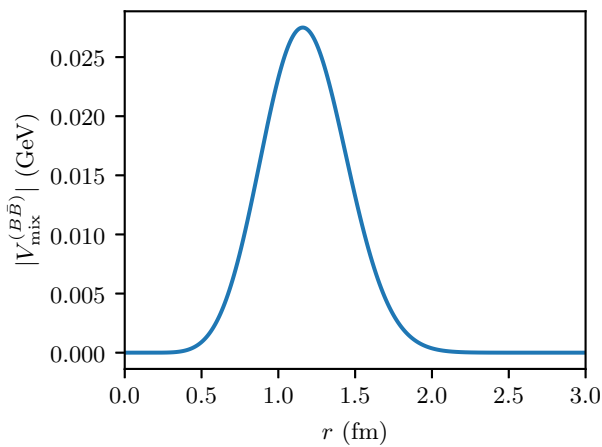}
\caption{Mixing potential between $b\bar{b}$ and $B \bar{B}$.\label{mixpot}}
\end{figure}

To make all this clear let us consider for example a system containing
$b\bar{b}$ and $B\bar{B}$. From Table~\ref{tablist} $r_\textup{c}^{(
B\bar{B})}=1.16$~fm. In the lattice calculation of Ref.~\cite{Bal05}, $r_\textup{c (lattice)}^{(B\bar{B})
}=1.25$~fm and $\Delta^{(B\bar{B})}_{\textup{(lattice)}}$ is close to $50$~MeV . Hence, we may expect
quite a similar value for $\Delta^{(B\bar{B})}$. As for $\rho$ we compare the
mixing angle between the ground and excited light field configurations
associated to $b\bar{b}$ and $B\bar{B}$ \cite{Bru20}:
\begin{equation}
\theta(r)=\frac{1}{2}\arctan\biggl(\frac{2V_\textup{mix}^{(
B\bar{B})}(r)}{T^{(B\bar{B})
}-V_{\text{C}}(r)}\biggr)
\end{equation}
to the one extracted from lattice, see Figure~15 in \cite{Bal05}. More
concretely, by using
\begin{subequations}\begin{align}
\Delta^{(B\bar{B})} &  =55\text{ MeV}\\
\rho &  =0.3\text{ fm}%
\end{align}\end{subequations}
we obtain the angle and mixing potential drawn in Figures~\ref{mixang} and \ref{mixpot} respectively.

It is worth remarking that the mixing is only effective in an interval around
$r_\textup{c}^{(B\bar{B})}$ determined by the value of $\rho$ and that the sign of the mixing potential
has no effect on the results that follow.

\begin{figure}
\centering
\input{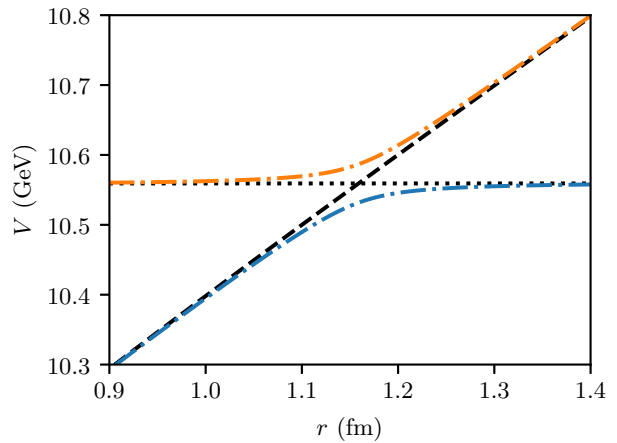}
\caption{Static energies. Dashed line: $b\bar{b}$ (Cornell). Dotted line: $B\bar{B}$ threshold.
Dash-dotted lines: $r$-dependent eigenvalues of the diabatic potential matrix \eqref{2dmat}.\label{dpots}}
\end{figure}

The diabatic potential matrix reads%
\begin{equation}%
\begin{pmatrix}
V_{\text{C}}(r) & V_{\textup{mix}}^{(B\bar{B})}(r)\\
V_{\textup{mix}}^{(B\bar{B})}(r) & m_{B}+m_{\bar{B}}%
\end{pmatrix}
\label{2dmat}
\end{equation}
and its eigenvalues, given by%
\begin{multline}
V_{\pm}(r)=\frac{V_{\text{C}}(r)+(m_{B}+m_{\bar{B}%
})}{2}\\
\pm\sqrt{\biggl(\frac{V_{\text{C}}(r)-(m_{B}%
+m_{\bar{B}})}{2}\biggr)^{2}+(V_{\textup{mix}%
}^{(B\bar{B})}(r))^{2}}%
\end{multline}
and represented in Figure~\ref{dpots}, should be compared to the static energies for
$b\bar{b}$ in the presence of $B\bar{B}$ calculated in lattice, see
Figures~13 and 14 in \cite{Bal05}. (Let us realize that the comparison has to be
more qualitative than quantitative since the values of the parameters in the
lattice depend on the chosen lattice spacing.)

\begin{figure}
\centering
\input{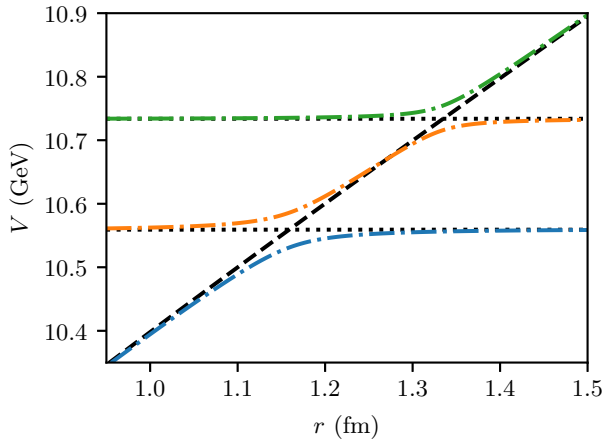}
\caption{Static energies. Dashed line: $b\bar{b}$ (Cornell). Dotted lines: $B\bar{B}$ and $B_{s}\bar{B}_{s}$ thresholds.
Dash-dotted lines: $r$-dependent eigenvalues of the diabatic potential matrix \eqref{3dmat}.\label{dpotsmany}}
\end{figure}

The extension to a system containing $b\bar{b}$, $B\bar{B}$ and
$B_{s}\bar{B}_{s}$ is straightforward. The diabatic potential matrix is
now%
\begin{equation}
\begin{pmatrix}
V_{\text{C}}(r) & V_{\textup{mix}}^{(B\bar{B})}(r) & V_{\textup{mix}%
}^{(B_{s}\bar{B}_{s})}(r)\\
V_{\textup{mix}}^{(B\bar{B})}(r) & m_{B}+m_{\bar{B}} & 0\\
V_{\textup{mix}}^{(B_{s}\bar{B}_{s})}(r) & 0 & m_{B_{s}}+m_{\bar
{B}_{s}}.
\end{pmatrix}
\label{3dmat}
\end{equation}
By using%
\begin{equation}
\Delta^{(B_{s}\bar{B}_{s})}=\frac{55}{\sqrt{2}}\text{ MeV}%
\end{equation}
the resulting eigenvalues, plotted in Figure~\ref{dpotsmany}, should be compared to the
educated guess of the static energies for $\ b\bar{b}$ in the presence of
$B\bar{B}$ and $B_{s}\bar{B}_{s}$ done in \cite{Bal05} (Figure 22) and
to the lattice calculation performed in \cite{Bul19}.

\section{State description\label{SIII}}

Any hidden-bottom meson state, characterized by the quantum numbers $J^{PC}$,
with mass below all possible open-bottom meson-meson thresholds with the same quantum
numbers is stable under decay into open-bottom meson-meson channels. Hence it
corresponds to a bound state solution of the diabatic multichannel
Schr\"{o}dinger equation. On the other hand any $J^{PC}$ hidden-bottom meson
state with mass above a possible open-bottom meson-meson threshold with the same quantum
numbers is unstable under decay into open-bottom meson-meson channels and
corresponds to a scattering solution of the diabatic multichannel
Schr\"{o}dinger equation.

From a technical point of view the extraction of the values of physical
observables from a (normalizable) bound state wavefunction is
straightforward. In contrast, for a scattering wavefunction it requires the
development of a dedicated formalism \cite{Scat}. Taking into account that the difference in the wave
functions is associated to the presence of open meson-meson components in the
scattering case, which are asymptotically free, one can approach a scattering
resonance solution through a two-step procedure. In the first step one solves
a bound state problem incorporating only the closed meson-meson components; in
the second step, one generates a resonance through the coupling of the bound
state solution to the open meson-meson components. This coupling allows for
the calculation of the mass of the resonance through mass corrections to the
bound state mass, and for the evaluation of its width. Notice that this
procedure is completely nonperturbative. One should keep in mind though the
lack of consistency of this bound state based approximation when there is some
open threshold giving rise to a mass correction to the bound state that makes
the threshold to close with respect to the resulting resonance. This was for example
the case of the hidden-charm meson $\psi(4040)$, see \cite{Bru21}.
Then, only a direct scattering solution of the multichannel
Schr\"{o}dinger equation can provide a trustable description.

\subsection{Bound states}

In order to calculate bound states a finite number of closed meson-meson
thresholds is considered. This is justified because in general
for a given bound state the probability of meson-meson components
corresponding to thresholds far above the mass of the bound
state is expected to be negligible. From
our mixing potential, we can estimate that this is the case for any threshold
being at least $200$~MeV above the mass of the bound state.

The technical procedure to calculate bound states has been detailed elsewhere,
see Sec.~IV~F and Appendices C and D in \cite{Bru20}. Let us only recall
here that in order to avoid possible multiple countings of the same bound
state when different sets of closed meson-meson thresholds are considered we
assume a one-to-one correspondence with the bound states of $b\bar{b}$
from the Cornell potential. Hence, each $b\bar{b}$ bound state is the
seed of only one bound state, the one obtained when the chosen set of closed
thresholds is maximal in the sense of containing the maximum possible number
of them. This assumption has proved to work for hidden-charm mesons
\cite{Bru20,Bru21}, and we shall show it also does for hidden-bottom ones.


Henceforth we center on spin-triplet hidden-bottom mesons with $I=0$ and
$J^{PC}=(0,1,2)^{++}$ and $1^{--}$ for which there are spectral
data available up to $11.0$ GeV. We restrict our study to bound states with
masses up to about $10.8$ GeV, two hundreed of MeV below the first $1^{--}$
$S$-wave meson-meson thresholds. There are several reasons for this. First,
it is known the $1^{1}P_{1}$ state $B_{1}(5721)$ but not the
corresponding $1^{3}P_{1}$ state $B_{1}(?)$ with an expected
similar mass. Hence, the threshold $B_{1}\bar{B}$ is only partially
known. Second, $B_{1}$ has a non-negligible width, $27.5\pm3.4$ MeV, and
$B_{1}(?)$ would presumably have a much wider state (actually,
this may be preventing its experimental detection). Hence, threshold width
effects should be properly incorporated.\ Third, the lowest lying bottomonium
hybrid $b\bar{b}g$ ($g:$ gluon), which could mix with $b\bar{b}$, is
predicted to have a mass about $10.9$ GeV, see \cite{Bru19} and references therein.


The possible values of $l_{b\bar{b}}$ $(s_{b\bar{b}%
}=1)$ and $(l_{M_{1}^{(i)}\bar{M}_{2}^{(i)}},s_{M_{1}%
^{(i)}\bar{M}_{2}^{(i)}})$ coupling to a given $J^{PC}$ are listed in
Table~\ref{quanum}, where the common notation $B_{(s)}$ to
refer to bottomed and bottomed strange mesons, and the shorthand
$B_{(s)}\bar{B}_{(s)}^{\ast}$ to denote the $C$-parity
eigenstate, are used.

\begin{table}
\caption{\label{quanum}Possible values of $l_{b\bar{b}}$ $(s_{b\bar{b}}=1)$ and $(l_{M_1^{(i)}\bar{M}^{(i)}_2}, s_{M_1^{(i)}\bar{M}^{(i)}_2})$ for given values of $J^{PC}$. A missing entry means that the particular meson-meson configuration cannot form a state with the corresponding quantum numbers.}
\begin{ruledtabular}
\begin{tabular}{ccccc}
$J^{PC}$	& $b\bar{b}$	& $B_{(s)} \bar{B}_{(s)}$	& $B_{(s)} \bar{B}_{(s)}^*$		& $B_{(s)}^{*} \bar{B}_{(s)}^*$ 	\\
\hline
$0^{++}$	& $1$		& $(0,0)$					& 								& (0,0), (2,2)						\\
$1^{++}$	& $1$		& 						& $(0,1),(2,1)$						& $(2,2)$							\\
$2^{++}$	& $1, 3$		& $(2,0)$					& $(2,1)$							& $(0,2),(2,0),(2,2),(4,2)$ 				\\
$1^{--}$	& $0, 2$		& $(1,0)$					& $(1,1)$							& $(1,0), (1,2), (3,2)$					\\
\end{tabular}
\end{ruledtabular}
\end{table}

The calculated spectrum of bound states is shown in Table~\ref{specomp}.

\begin{table*}
\caption{Calculated masses, $b\bar{b}$ and meson-meson
component probabilities, for $J^{PC}$ bottomoniumlike bound state solutions.
Vanishing and negligible (i.e., inferior to $1\%$) probabilities are not displayed.\label{specomp}}
\begin{ruledtabular}
\begin{tabular}{ddddddddd}
J^{PC}	& \text{Mass (MeV)}	& b\bar{b}	& B \bar{B}	& B \bar{B}^*	& B^* \bar{B}^*	& B_s \bar{B}_s 	& B_s \bar{B}_s^*	& B_s^* \bar{B}_s^*	\\
\hline
0^{++}	& 9900.7			& 100 \%		& 		& 		& 			& 		& 			& 			\\
		& 10254.1			& 100 \%		& 		& 		& 			& 		& 			& 			\\
		& 10530.2			& 91 \%		& 8 \%	& 		& 1 \%		& 		& 			& 			\\
		& 10778.1			& 98 \%		& 		& 		& 			& 		& 			& 2 \%		\\
1^{++}	& 9900.7			& 100 \%		& 		& 		& 			& 		& 			& 			\\
		& 10254.2			& 100 \%		& 		& 		& 			& 		& 			& 			\\
		& 10532.1			& 97 \%		& 		& 3 \%	& 			& 		& 			& 			\\
		& 10775.1			& 75 \%		& 		& 		& 			& 		& 24 \%		& 1 \%		\\
2^{++}	&  9900.7			& (100, 0) \%	& 		& 		& 			& 		& 			& 			\\
		& 10253.9			& (100, 0) \%	& 		& 		& 			& 		& 			& 			\\
		& 10340.9			& (0, 100) \%	& 		& 		& 			& 		& 			& 			\\
		& 10527.7			& (92, 2) \%	& 3 \%	& 1 \%	& 2 \%		& 		& 			& 			\\
		& 10592.7			& (2, 91) \%	& 		& 3 \%	& 4 \%		& 		& 			& 			\\
		& 10776.2			& (91, 1) \%	& 		& 		& 			& 		& 4 \%		& 4 \%		\\
1^{--}	& 9401.2			& (100, 0) \%	& 		& 		& 			& 		& 			& 			\\
		& 9993.8			& (100, 0) \%	& 		& 		& 			& 		& 			& 			\\
		& 10150.3			& (0, 100) \%	& 		& 		& 			& 		& 			& 			\\
		& 10337.2			& (100, 0) \%	& 		& 		& 			& 		& 			& 			\\
		& 10439.4			& (0, 99) \%	& 1 \%	& 		& 			& 		& 			& 			\\
		& 10598.8			& (70, 3) \%	& 		& 21 \%	& 6 \%		&		& 			& 			\\
		& 10691.3			& (0, 98) \%	& 		& 		& 			& 1 \%	&			& 1 \%		\\
\end{tabular}
\end{ruledtabular}
\end{table*}

A glance at the table and its comparison with Table~\ref{tabcor} makes clear that i) all
bound states have a dominant $b\bar{b}$ component, with more than $90\%$
probability in most cases, ii) closed meson-meson thresholds give rise to
attraction, iii) the attractive effect on the mass is quantitatively modest,
with mass reductions of $16$ MeV or less with respect to the $b\bar{b}$
masses obtained from the Cornell potential. These results are in line with the
reasonable mass description of known experimental resonances provided by the
Cornell potential model.

For a detailed comparison to data we have to take into account that our
Cornell potential does not contain spin-dependent terms. Then, for pure $(nl)$
$b\bar{b}$ states the calculated masses have to be compared to the $(nl)$
experimental centroids; in the other cases, where meson-meson components are
present, since they are specific for any set of $J^{PC}$ quantum numbers, the
comparison has to be done with the experimental candidates with the same
$J^{PC}$. Taking this into consideration all known $J^{PC}=(
0,1,2)^{++}$ and $1^{--}$ experimental resonances below $10.8$ GeV can
be assigned to bound states with the same location (below or between) with
respect to the meson-meson thresholds. Thus, we see that the calculated mass
for the $(0,1,2)^{++}$ ground states, which are $100\%$ Cornell $(
1P)$ $b\bar{b}$ states, coincides with the experimental mass
centroid from $1P_{J}$ states at $9899.9\pm0.6$~MeV. Actually, this
coincidence has been required to fix the bottom quark mass. As for the first
excited $(0,1,2)^{++}$ states, which are $100\%$ Cornell $(2P)$
$b\bar{b}$ states, the calculated mass is very close to the experimental
mass centroid from $2P_{J}$ states at $10260.2\pm0.7$~MeV. The only additional
$J^{++}$ pure Cornell state is the second excitation of $2^{++}$, assigned to
the $(1F_{2})$ $b\bar{b}$ state with a predicted mass of
about $10340$~MeV.

The second excited $(0,1)^{++}$ and the third excited $2^{++}$ states are
predicted to contain more than a $90\%$ of $(3P)$
$b\bar{b}$ and less than a $10\%$ of meson-meson components. For $1^{++}$
and $2^{++}$ the calculated masses compare well with existing data (for
$0^{++}$ there is no PDG data). Indeed the measured masses of $\chi
_{b1}(3P)$, $10513.42\pm0.41\pm0.53$~MeV, and $\chi_{b2}(
3P)$, $10524.02\pm0.57\pm0.53$~MeV, differ from the calculated values
by less than $30$ MeV. It is worth mentioning that the presence of meson-meson
components makes the calculated masses to be $12$ MeV closer to data than the
corresponding Cornell masses suggesting that a renaming of these resonances as
$\chi_{b1}(10513)$ and $\chi_{b2}(10524)$ might
be in order.

From the point of view of its meson-meson composition the most interesting
$J^{++}$ case is the third excited state of $1^{++}$ with a significant $24\%$
of $B_{s}\bar{B}_{s}^{\ast}$. This significant percentage has to do with
the immediate vicinity of the Cornell $(4P)$ $b\bar{b}$
state and the $B_{s}\bar{B}_{s}^{\ast}$ threshold, both located at about
$10782.2$~MeV. Notice though that the mass shift due to $B_{s}\bar{B}%
_{s}^{\ast}$ is only $7$~MeV with respect to the Cornell $(4P)$
$b\bar{b}$ mass. More importantly, as $B_{s}\bar{B}_{s}^{\ast}$ can
naturally decay strongly into $\Upsilon(1S)\phi$ through quark
exchange this could be a possible discovery channel.


For $1^{--}$ the ground an the first three excited states are predicted to be
$100\%$ the Cornell $(1S,2S,1D,3S)$ $b\bar{b}$ states
respectively. For the ground state $(1S)$ the difference
between the experimental centroid (from $\eta_{b}(1S)$ and
$\Upsilon(1S))$ at $9445.0\pm0.7$~MeV and the calculated mass
is $45$~MeV, significantly higher than in any other case. This could be
indicating the presence of more relevant relativistic effects in the $1S$
state. Indeed, for the first excited state $(2S)$ the
difference between the $(\eta_{b}(2S),\Upsilon(
2S))$ mass centroid at $10017.20\pm1.8$~MeV and the calculated
value gets reduced to $23$~MeV. For the $(1D)$ and $(
3S)$ states the lack of data prevents the evaluation of the mass
centroids for comparison. Instead, we can check that the calculated mass for
$(1D)$ is pretty close to the measured mass of $\Upsilon
_{2}(1D)$, $10163.7\pm1.4$~MeV, and that the calculated mass of
$(3S)$ is lower than the measured mass of $\Upsilon(
3S)$, $10355.2\pm0.5$~MeV, as should be expected.

All the higher excited states contain meson-meson components. However, only
for the fifth excited state, which contains a dominant $(70\%)$
Cornell $(4S)$ $b\bar{b}$ component, we predict a
significant meson-meson probability $(21\%\text{ of }B\bar
{B}^{\ast})$, due to the vicinity of the $(4S)$
$b\bar{b}$ state and the $B\bar{B}^{\ast}$ threshold. This suggests
that for the corresponding experimental resonance the label
$\Upsilon(10580)$ should be preferred to the PDG alternative
$\Upsilon(4S)$.

\begin{figure}
\centering
\input{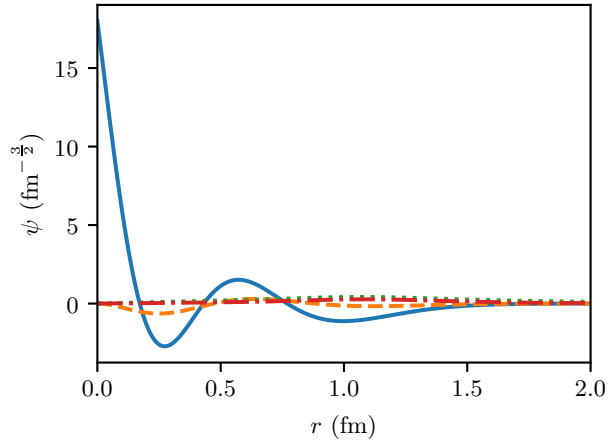}
\caption{Radial wavefunction of the fifth excited $1^{--}$ state. $b\bar{b}(4 S)$, $b\bar{b}(3 D)$, $B \bar{B}^{\ast}(l_{B \bar{B}^{\ast}}=1)$, and $B^{\ast} \bar{B}^{\ast}(l_{B^{\ast} \bar{B}^{\ast}}=1)$ components are drawn with a solid, dashed, dotted, and dash-dotted line respectively.\label{upsi10580}}
\end{figure}

It is illustrative to plot the radial wavefunction of this state for the several
components, see Figure~\ref{upsi10580}.

We see that the presence of meson-meson components makes the radial wave
function to extend to larger distance than the Cornell $(4S)$
$b\bar{b}$ one, and correlated with this there is a loss of probability
density at the origin $(r=0)$ as compared to the Cornell case.
This could explain the discrepancies observed between the calculated leptonic
width ratios in the Cornell model and data. More concretely, the $1^{--}$
leptonic width ratios are calculated from \cite{Eic08}
\begin{equation}
\frac{\Gamma(\Upsilon^{(n_{1})}\rightarrow e^{+}%
e^{-})}{\Gamma(\Upsilon^{(n_{2})}\rightarrow
e^{+}e^{-})}= \biggl\lvert\frac{R_{\Upsilon^{(n_{1})}}(0)}{R_{\Upsilon^{(n_{2})}}(0)}\biggr\rvert^{2}\frac{m_{\Upsilon^{(n_{2})}}^{2}}%
{m_{\Upsilon^{(n_{1})}}^{2}}%
\end{equation}
where $R_{\Upsilon^{(n)}}(0)$ stands for the
radial wavefunction at the origin and $m_{\Upsilon^{(n)}}$ for
the mass of the $\Upsilon$ state containing a $(nS)$
$b\bar{b}$ component. The calculated values for these ratios and their
comparison to data are given in Table~\ref{VI} (the use of the experimental masses instead
of the calculated ones would not make any difference).

\begin{table}
\caption{Calculated leptonic width ratios from the Cornell model and the
diabatic approach, as compared to data from \cite{PDG20}.\label{VI}}
\begin{ruledtabular}
\begin{tabular}{cccc}
Leptonic Width Ratio & Cornell & Experiment & Diabatic\\
\hline
$\frac{\Gamma(\Upsilon(2s)\rightarrow e^{+}e^{-})
}{\Gamma(\Upsilon(1s)\rightarrow e^{+}e^{-})}$ &
$0.36$ & $0.456\pm0.14$ & $0.36$\\
$\frac{\Gamma(\Upsilon(3s)\rightarrow e^{+}e^{-})
}{\Gamma(\Upsilon(1s)\rightarrow e^{+}e^{-})}$ &
$0.25$ & $0.33\pm0.1$ & $0.25$\\
$\frac{\Gamma(\Upsilon(10580)\rightarrow e^{+}%
e^{-})}{\Gamma(\Upsilon(1s)\rightarrow
e^{+}e^{-})}$ & $0.21$ & $0.20\pm0.02$ & $0.14$\\
$\frac{\Gamma(\Upsilon(3s)\rightarrow e^{+}e^{-})
}{\Gamma(\Upsilon(2s)\rightarrow e^{+}e^{-})}$ &
$0.70$ & $0.72\pm0.03$ & $0.70$\\
$\frac{\Gamma(\Upsilon(10580)\rightarrow e^{+}%
e^{-})}{\Gamma(\Upsilon(2s)\rightarrow
e^{+}e^{-})}$ & $0.58$ & $0.44\pm0.06$ & $0.39$\\\
$\frac{\Gamma(\Upsilon(10580)\rightarrow e^{+}%
e^{-})}{\Gamma(\Upsilon(3s)\rightarrow
e^{+}e^{-})}$ & $0.82$ & $0.61\pm0.08$ & $0.56$%
\end{tabular}
\end{ruledtabular}
\end{table}

A look at the table makes clear that i) the ratios involving
$\Gamma(\Upsilon(1S)\rightarrow e^{+}e^{-})$,
are defficiently described by both the Cornell model (with the exception of
$\frac{\Gamma(\Upsilon(10580)\rightarrow e^{+}%
e^{-})}{\Gamma(\Upsilon(1s)\rightarrow
e^{+}e^{-})})$ and the diabatic approach, ii) the diabatic values for
these ratios can be put in accord with data through multiplication by a common
factor of $\simeq1.3$, iii) all the ratios not involving $\Gamma(
\Upsilon(1S)\rightarrow e^{+}e^{-})$ are well described
by the diabatic approach whereas the Cornell model is far from data except
$\frac{\Gamma(\Upsilon(3s)\rightarrow e^{+}e^{-})
}{\Gamma(\Upsilon(2s)\rightarrow e^{+}e^{-})}$
for which there is no difference in the calculated wavefunctions with both approximations.

These results suggest that the failure of the diabatic approach regarding the
ratios involving $\Gamma(\Upsilon(1S)\rightarrow
e^{+}e^{-})$ may have to do with the presence of relativistic
corrections in $\Upsilon(1S)$ making its radial wavefunction
at the origin to decrease a $14\%$.

Therefore, we may tentatively conclude that data from leptonic widths can be
taken as an indication of the meson-meson compositeness of $\Upsilon(
10580)$. In this regard, it is also interesting to add that it is the
$b\bar{b}$ $(4S)$-$B\bar{B}^{\ast}$ interaction the
main physical mechanism underlying the explanation of the leptonic widths. The
small $D$ mixing, $3\%$ of $b\bar{b}$ $(3D)$, which is
mainly induced through the $S$ and $D$ coupling to $B^{\ast}\bar{B}%
^{\ast}$, plays a marginal quantitative role. Moreover, other sources of $S$-$D$
mixing such as a direct tensor interaction within the Cornell potential,
should also have a quite limited importance in order to preserve the accurate
leptonic width description. This is in contrast to other explanations in the
literature based on a significant $S$-$D$ mixing, see for instance \cite{Bad09}.


The calculated sixth excited state has a predicted mass close to $10700$~MeV
and it is very dominantly a $b\bar{b}$ $(3D)$ state. It
could be possibly assigned to the not well established $\Upsilon(10753)$ with
a measured mass of $10752.7\pm5.9_{-1.1}^{+0.7}$~MeV. However, the discovery
channel $\Upsilon(nS)\pi^{+}\pi^{-}$ with $n=1,2,3$ is not expected to be a
dominant channel for a $(3D)$ state suggesting that some mixing
with the $b\bar{b}$ $(4S)$ state is lost. This can be due
to the fact that the $B\bar{B}^{\ast}$, and particularly the $B^{\ast
}\bar{B}^{\ast}$ thresholds, which according to our previous discussion
can give rise to this mixing, are not taken into account in our bound state
approach for the sixth excited state since they are open meson-meson
channels. Indeed, we shall show later on that this excitation has a prominent
width to $B^{\ast}\bar{B}^{\ast}$. Hence, a significant $D$-$S$ mixing
through the coupling to $B^{\ast}\bar{B}^{\ast}$ could be present. The
theoretical description of this mixing would require a complete (scattering)
solution of the problem which is out of the scope of our current analysis.

\subsection{Mass corrections and widths}

Let us realize that for pure Cornell states, with masses below the first
meson-meson threshold, there are no open meson-meson channels. Hence, no
widths, and mass corrections being mostly limited to spin splittings which are
known to be quantitatively important for these states. Indeed, the derivation
of the form of the spin dependent potential terms from QCD, and a numerical
evaluation of the Cornell spin splittings for $b\bar{b}$ was carried out
fourty years ago \cite{Fei81}. We simply copy here those results for the
slightly different values we use for the parameters of the Cornell potential
hardly makes a difference. The corrected masses for the
pure Cornell states in Table~\ref{specomp} are given in Table~\ref{VII}.


\begin{table}
\caption{Spin splittings (in MeV) for pure Cornell spin triplet states. The corrected
masses (in MeV) and their comparison to the measured masses \cite{PDG20} of the
assigned mesons are also shown.\label{VII}}
\begin{ruledtabular}
\begin{tabular}{ccccc}
$J^{PC}(nl)$ & Splitting & Mass & Experiment & Meson\\
\hline
$1^{--}(1S)$ & $23.7$ & $9424.9$ & $9460.30\pm0.26$ &
$\Upsilon(1S)$\\
$0^{++}(1P)$ & $-35.8$ & $9864.9$ & $9859.44\pm0.42\pm0.31$ & $\chi
_{b0}(1P)$\\
$1^{++}(1P)$ & $-11.0$ & $9889.7$ & $9892.78\pm0.26\pm0.31$ & $\chi
_{b1}(1P)$\\
$2^{++}(1P)$ & $13.8$ & $9914.5$ & $9912.21\pm0.26\pm0.31$ & $\chi_{b2}(
1P)$\\
$1^{--}(2S)$ & $10.3$ & $10004.1$ & $10023.26\pm0.31$ &
$\Upsilon(2S)$\\
$1^{--}(1D)$ & $-3.3$ & $10147.0$ & $10163.7\pm1.4$ &
$\Upsilon_{2}(1D)$\\
$0^{++}(2P)$ & $-26.4$ & $10227.7$ & $10232.5\pm0.4\pm0.5$ & $\chi_{b0}(
2P)$\\
$1^{++}(2P)$ & $-8.2$ & $10246.0$ & $10255.46\pm0.22\pm0.50$ & $\chi
_{b1}(2P)$\\
$2^{++}(2P)$ & $10.2$ & $10264.1$ & $10268.65\pm0.22\pm0.50$ & $\chi
_{b2}(2P)$\\
$1^{--}(3S)$ & $7.8$ & $10345.0$ & $10355.2\pm0.5$ &
$\Upsilon(3S)$%
\end{tabular}
\end{ruledtabular}
\end{table}


We see that a very good mass description is obtained. The biggest difference
between the calculated mass and data is of $35$ MeV for $\Upsilon(
1S)$ which we may attribute, at least partially, to further
relativistic (kinetic energy) effects.


To proceed to a similar evaluation of spin splitting for states with
meson-meson components, spin dependent terms of the mixing and meson-meson
potentials should be taken into account as well. However, the complete lack of
knowledge of the spin dependence in the mixing potential prevents carrying out
this procedure. Instead, for states with mass above the first meson-meson
threshold, we can evaluate mass corrections and widths from the open
meson-meson thresholds neglected in the bound state calculation. The
nonperturbative method we follow for this evaluation has been explained
elsewhere, see \cite{Bru21} and references therein. Let us only recall here
that the physical effect of the coupling to the continuum is to dilute the
bound state through a band of stationary scattering states, giving rise to a
resonance. If we call $m_\textup{bs}$ the mass of the bound state, $M_{1}%
^{(j)}\bar{M}_{2}^{(j)}$ with $j=1,2\dots,n$ the corresponding open
meson-meson components, and $m_\textup{res}$ and $\Gamma$ the mass and width
respectively of the resulting resonance, then
\begin{equation}
m_\textup{res}-m_\textup{bs}=\sum_{j, k}\mathcal{P}\!\!\int \mathrm{d}E^{(j)}\mu^{(
j)}\frac{p^{(j)}}{m_\textup{res}-E^{(j)}%
} \bigl\lvert\mathcal{I}_{k}^{(j)}(p^{(j)})\bigr\rvert^{2}%
\end{equation}
and%
\begin{equation}
\frac{\Gamma}{2}=\sum_{j, k}\pi p^{(j)}\bigl\lvert\mathcal{I}_{k}^{(j)}(
p^{(j)})\bigr\rvert _{E^{(j)}=m_\textup{res}}^{2}%
\end{equation}
where $\mathcal{P}\int$\ stands for the Cauchy principal value integral,
$E^{(j)}$ for the meson-meson energy in the center of mass
frame (resonance at rest)
\begin{equation}
E^{(j)}=E_{M_{1}^{(j)}}+E_{\bar{M}_{2}^{(j)}}%
\end{equation}
with
\begin{subequations}
\begin{align}
E_{M_{1}^{(j)}} =& m_{M_{1}^{(j)}}+\frac{(p^{(j)})
^{2}}{2m_{M_{1}^{(j)}}} \\
E_{\bar{M}_{2}^{(j)}} =& m_{\bar{M}_{2}^{(j)}}+\frac{(p^{(j)})
^{2}}{2 m_{\bar{M}_{2}^{(j)}}}
\end{align}
\end{subequations}
where $p^{(j)}$ is the meson momentum, $\mu^{(j)
}$ for the reduced mass of $M_{1}^{(j)}$ and $\bar{M}_{2}^{(j)}$%
\begin{equation}
\mu^{(j)}\equiv \frac{m_{M_{1}^{(j)}} m_{\bar{M}_{2}^{(j)}}}{m_{M_{1}^{(j)}} 
+m_{\bar{M}_{2}^{(j)}}}
\end{equation}
and
\begin{multline}
\mathcal{I}_{k}^{(j)}(
p^{(j)})\equiv\sqrt{\frac{2}{\pi}}i^{-l_{k}^{(j)}}\\
\int \mathrm{d}r \, r^{2} j_{l_{k}^{(j)}}(p^{(j)}r)V_\textup{mix}^{(j)}(r)
\Bigl(\sum_{t}R_{t}^{(0)}(r)\Bigr)
\end{multline}
where $j_{l}$ is the spherical Bessel function.

It should be emphasized that our calculation of mass corrections and widths
does not introduce any new parameter, so the comparison of the masses and
widths of the resulting resonances with data may serve as a stringent test of
the mixing interaction.


The calculated masses of the resonances $m_\textup{res}$, as well as their
differences with the masses of the corresponding bound states, $m_\textup{res}%
-m_\textup{bs}$, are shown in Table~\ref{VIII}. For the sake of completeness we have also
included in the table non pure Cornell bound states whose masses are below the
first open-bottom meson-meson threshold and, consequently there is no mass correction.

\begin{table}
\caption{Calculated mass corrections, $m_\textup{res}-m_\textup{bs}$, and total masses,
$m_\textup{res}$, in MeV, for non pure Cornell $J^{PC}=(0,1,2)^{++}$
and $1^{--}$ states below 10.8 GeV. Measured masses from \cite{PDG20}, when
existing, corresponding to the meson type assignment in the last column (the
subscript referring to the dominant $b\bar{b}$ component), are also given
for comparison.\label{VIII}}
\begin{ruledtabular}
\begin{tabular}{ccccc}
$J^{PC}$ & $m_\textup{res}-m_\textup{bs}$ & $m_\textup{res}$ & Experiment & Meson\\
\hline
$0^{++}$ & $0$ & $10530.2$ &  & $(\chi_{b0})_{3P}$\\
& $7.7$ & $10785.8$ &  & $(\chi_{b0})_{4P}$\\
$1^{++}$ & $0$ & $10532.1$ & $10513.42\pm0.41\pm0.53$ & $\chi_{b1}(
10513)$\\
& $3.8$ & $10778.9$ &  & $(\chi_{b1})_{4P}$\\
$2^{++}$ & $0$ & $10527.7$ & $10524.02\pm0.57\pm0.53$ & $\chi_{b2}(
10524)$\\
& $-4.3$ & $10588.4$ &  & $(\chi_{b2})_{2F}$\\
& $6.1$ & $10782.3$ &  & $(\chi_{b2})_{4P}$\\
$1^{--}$ & $0$ & $10439.4$ &  & $(\Upsilon)_{2D}$\\
& $1.0$ & $10599.8$ & $10579.4\pm1.2$ & $\Upsilon(10580)$\\
& $5.7$ & $10697.0$ & $10752.7\pm5.9_{-1.1}^{+0.7}$ & $\Upsilon(
10753)$
\end{tabular}
\end{ruledtabular}
\end{table}

As can be checked the calculated masses are in good agreement (less
than $20$ MeV mass difference) with the few existing data, except for
$\Upsilon(10753)$. This may be indicating that the non
considered spin splittings are not quantitatively as important as for pure
Cornell states. On the other hand, the mass corrections due to open
thresholds are quantitatively small, of a few $(<10)$ MeV at
most. Altogether, these arguments make us confident about the predicted masses
for the not yet discovered resonances. Regarding $\Upsilon(10753)$ the deficient predicted mass may
be indicating the lack of a significant $D$-$S$ mixing which could arise from
the coupling to $B^{\ast}\bar{B}^{\ast}$ as well as from the
incorporation of spin dependent terms in the diabatic potential matrix.


As for the decay widths to open-bottom meson-meson channels, we have
summarized them in Table~\ref{maswid}.

\begin{table*}
\caption{Total masses, $m_\textup{res}$, and decay widths to open-bottom meson-meson channels, in MeV, of bottomoniumlike states above threshold. Available experimental widths from \cite{PDG20} are quoted for comparison.\label{maswid}}
\begin{ruledtabular}
\begin{tabular}{dddddddc}
J^{PC}	& m_\textup{res}		& \Gamma_{B \bar{B}}	& \Gamma_{B \bar{B}^*}	& \Gamma_{B^* \bar{B}^*}	& \Gamma_{B_s \bar{B}_s}	& \Gamma_{\textup{tot}}^{\textup{Theor}}	& $\Gamma_{\textup{tot}}^{\textup{Expt}}$\\
\hline
0^{++}	& 10785.8		& 1.6				& 				& 5.3					& 0.7				& 7.6								&								\\
1^{++}	& 10778.9		&				& 0.2				& 1.7					& 				& 1.9		 						&								\\
2^{++}	& 10588.4		& 4.3				&				&					&				& 4.3								&								\\
		& 10782.3		& 5.4				& 1.5				& 21.0				& 10.4			& 38.3							&								\\
1^{--}	& 10599.8		& 21.9			& 				& 					& 				& 21.9	 						& $20.5 \pm 2.5$					\\
		& 10697.0		& 2.0				& 1.0				& 38.0				& 				& 41.0							& $36^{+17.6 + 3.9}_{-11.3-3.3}$		\\
\end{tabular}
\end{ruledtabular}
\end{table*}

It is remarkable that the calculated values are in very good agreement with
the few existing data. So, for $\Upsilon(10580)$ and
$\Upsilon(10753)$ the predicted values of the total widths are
fully compatible with data. For $\Upsilon(10580)$ there is only
one decay channel $B\bar{B}$ contributing to the width. Experimentally
this channel saturates with more than $96\%$ the total width. For the not well
established $\Upsilon(10753)$ we predict a very dominant decay
into $B^{\ast}\bar{B}^{\ast}$ which could help to guide new experimental
searches. It is also worth to emphasize the pretty small values of the
widths predicted for some not yet discovered $(0,1,2)^{++}$ resonances, in
particular for the third excited $1^{++}$ state with a value smaller than $2$
MeV. Although we should add to these predictions some uncertainty we consider
them encouraging for experimental analyses. In the $1^{++}$ case, its small
width (to $B^{\ast}\bar{B}^{\ast}$ and $B^{\ast}\bar{B}^{\ast}$), and
its $B_{s}\bar{B}_{s}^{\ast}$ content (see Table~\ref{specomp}) pointing
out to a significant $\Upsilon(1S)\phi$ decay, can help to its experimental discovery.


It should be mentioned that the effect of meson-meson thresholds in the
hidden-bottom spectrum has been previously studied in the literature, see for
instance \cite{vanB83,Ono84,Fer14} and references therein. Quite generally all
these studies make use of the $^{3}\!P_{0}$ quark pair creation mechanism to mix
the $b\bar{b}$ and meson-meson components. This $^{3}\!P_{0}$ model lacks
direct justification from QCD. Instead, we use a mixing potential whose form
is directly based on lattice results for the static light field energies for
$b\bar{b}$ in the presence of meson-meson configurations. This gives rise
to an utterly different mixing description which is enhanced around the
crossing points of the Cornell potential with the meson-meson thresholds.
Moreover, in most of these previous studies a perturbative hadron-loop
treatment has been followed to calculate mass corrections and widths from
meson-meson thresholds. This is conceptually questionable since the
perturbative series is divergent as shown in \cite{Bru21}. Much more related
to our work is a quite recent analysis \cite{Bic21} where the authors proceed
to a numerical parametrization of the diabatic potential matrix from lattice
data and to the evaluation from it of the scattering solutions of the
Schr\"{o}dinger equation. However, due to the formidable difficulty of the
problem only $b\bar{b}$ $S$-waves and a reduced number of thresholds masses (at
most two) have been considered until now, what reduces considerably its
predictive power.

\section{Summary\label{SIV}}

We have applied the diabatic approach, which allows for a QCD based analysis
of the heavy-quark meson spectra, to the description of $I=0$, $J^{PC}=(
0,1,2)^{++}$ and $1^{--}$ hidden-bottom mesons with masses below $10.8$
GeV. More precisely, we have solved the Schr\"{o}dinger equation for
$b\bar{b}$ and meson-meson components with the same $J^{PC}$ quantum
numbers. The form of the diabatic potential matrix entering in this equation
has been derived from current lattice data on the energy of static $b$ and
$\bar{b}$ sources, when $b\bar{b}$ mixes with $B\bar{B}$ and
$B_{s} \bar{B}_{s}$ configurations. For practical purposes, we have
followed a bound state based approximation to describe resonances. From it mass corrections
and widths have been properly incorporated. A good description of masses,
decay widths to open-bottom meson-meson channels, and (for $1^{--}$) leptonic
widths of known resonances has been obtained. Of particular interest is the
prediction of a significantly lower than $1$ $b\bar{b}$ probability in
$\Upsilon(10580)$ which allows for an accurate description of its leptonic
width ratios with the $\Upsilon((2,3)S)$ states. Concerning the not yet
discovered resonances it is noteworthy the prediction of a narrow $\chi
_{b1}(10779)$, possibly with a significant decay into
$\Upsilon(1S)\phi$ as a consequence of its $B_{s}\bar
{B}_{s}^{\ast}$ content.

These results for bottomoniumlike mesons give additional support to the
diabatic approach in QCD as an appropriate framework for a unified and
complete nonperturbative description of conventional and unconventional
heavy-quark mesons. They make also clear that the presence of meson-meson
thresholds introduces a flavor spectral dependence so that a flavor
independent partner correspondence between the hidden-bottom and the
hidden-charm mesons is limited to the conventional bottomonium, pure
$b\bar{b}$, and charmonium, pure $c\bar{c}$, states.

\begin{acknowledgments}
This work has been supported by \foreignlanguage{spanish}{Ministerio de Economía, Industria y Competitividad} of Spain and European Regional Development Fund Grant No.~FPA2016-77177-C2-1-P, by EU Horizon 2020 Grant No.~824093 (STRONG-2020), and by \foreignlanguage{spanish}{Ministerio de Ciencia e Innovación} and \foreignlanguage{spanish}{Agencia Estatal de Investigación} of Spain and European Regional Development Fund Grant PID2019-105439 GB-C21. R.~B. acknowledges a FPI fellowship from \foreignlanguage{spanish}{Ministerio de Ciencia, Innovacíon y Universidades} of Spain under Grant No. BES-2017-079860.
\end{acknowledgments}

\appendix*

\section{Degenerate thresholds}

Let us consider for instance the (almost) degenerate $B^{+}B^{-}$ and
$B^{0}\bar{B}^{0}$ thresholds.

In the diabatic formalism the potential matrix for a physical system
containing $b\bar{b}$, $B^{+}B^{-}$ and $B^{0}\bar{B}%
^{0}$ components reads%
\begin{equation}
\begin{pmatrix}
V_{\text{C}}(r) & V_{\textup{mix}}^{(1)}(r) & V_{\textup{mix}}^{(2)}(r)\\
V_{\textup{mix}}^{(1)}(r) & T_{B^{+}B^{-}} & 0\\
V_{\textup{mix}}^{(2)}(r) & 0 & T_{B^{0}\bar{B}^{0}}%
\end{pmatrix}
\end{equation}
with%
\begin{subequations}\begin{align}
V_{\textup{mix}}^{(1)}(r)  &  =\bra{ \zeta_{b\bar{b}%
}} H_\textup{static}^\textup{lf}(\bm{r})\ket{ \zeta
_{B^{+}B^{-}}} \\
V_{\textup{mix}}^{(2)}(r)  &  =\bra{ \zeta_{b\bar{b}%
}} H_\textup{static}^\textup{lf}(\bm{r})\ket{ \zeta
_{B^{0}\bar{B}^{0}}} \\
V_{\text{C}}(r)  &  =\bra{ \zeta_{b\bar{b}}}
H_\textup{static}^\textup{lf}(\bm{r})\ket{ \zeta_{b\bar{b}%
}} \\
T_{B^{+}B^{-}}  &  =\bra{ \zeta_{B^{+}B^{-}%
}} H_\textup{static}^\textup{lf}(\bm{r})\ket{ \zeta
_{B^{+}B^{-}}} \\
T_{B^{0}\bar{B}^{0-}}  &  =\bra{ \zeta_{B^{0}\bar{B}^{0}%
}} H_\textup{static}^\textup{lf}(\bm{r})\ket{ \zeta
_{B^{0}\bar{B}^{0}}}
\end{align}\end{subequations}
where $H_\textup{static}^\textup{lf}(\bm{r})$ denotes the hamiltonian for the
light fields (gluons and light quarks), and $\ket{ \zeta
_{b\bar{b}}} $, $\ket{ \zeta_{B^{+}B^{-}%
}} $, $\ket{ \zeta_{B^{0}\bar{B}^{0}}}
$, the ground and excited light field configurations associated to the
$b\bar{b}$, $B^{+}B^{-}$ and $B^{0}\bar{B}^{0}$
components respectively \cite{Bru20}.


If we neglect the small mass difference between $T_{B^{+}B^{-}}$and
$T_{B^{0}\bar{B}^{0}}$, assume $V_{\textup{mix}}^{(1)}%
(r)=V_{\textup{mix}}^{(2)}(r)\equiv V_{\textup{mix}}(r)$, and define the light field configurations
\begin{subequations}\begin{align}
\ket{ \zeta_{(B\bar{B})_{0}}}  &
=\frac{1}{\sqrt{2}}\ket{ \zeta_{B^{+}B^{-}}}
-\frac{1}{\sqrt{2}}\ket{ \zeta_{B^{0}\bar{B}^{0}}}
\\
\ket{ \zeta_{(B\bar{B})_{1}}}  &
=\frac{1}{\sqrt{2}}\ket{ \zeta_{B^{+}B^{-}}}
+\frac{1}{\sqrt{2}}\ket{ \zeta_{B^{0}\bar{B}^{0}}}
\end{align}\end{subequations}
then the expression of the diabatic potential matrix in terms of these light
field configurations reads
\begin{equation}
\begin{pmatrix}
V_{\text{C}}(r) & \sqrt{2}V_{\textup{mix}}(r) & 0\\
\sqrt{2}V_{\textup{mix}}(r) & T_{B\bar{B}} & 0\\
0 & 0 & T_{B\bar{B}}
\end{pmatrix}
\end{equation}
where $T_{B\bar{B}}=T_{B^{+}B^{-}}=T_{B^{0}\bar
{B}^{0}}$ and
\begin{subequations}\begin{align}
\bra{ \zeta_{b\bar{b}%
}} H_\textup{static}^\textup{lf}(\bm{r})\ket{ \zeta_{(B\bar{B})_{0}}} &= \sqrt{2}V_{\textup{mix}}(r) \\
\bra{ \zeta_{b\bar{b}}} H_\textup{static}^\textup{lf}(\bm{r})\ket{ \zeta_{(B\bar{B})_{1}}} &= 0
\end{align}\end{subequations}

Hence, as the light field configuration $\ket{ \zeta_{B\bar{B}%
}} _{1}$ does not couple to $\ket{ \zeta_{b\bar{b}%
}} $ the physical system is equivalent to two decoupled
subsystems, one containing $b\bar{b}$ and $(B\bar{B})
_{0}$ components and the other containing the $(B\bar{B})
_{1}$ component. Actually, $(B\bar{B})_{0}$ and $(
B\bar{B})_{1}$ correspond to an isosinglet and to the neutral
projection of an isotriplet respectively. Therefore, the lack of mixing
between $\ket{ \zeta_{b\bar{b}}} $, associated to
the $b\bar{b}$ isosinglet, and $\ket{ \zeta_{B\bar{B}%
}} _{1}$ is just a consequence of the isospin conservation in
strong interactions.

These results make clear that the effect of two degenerate thresholds such as
$B^{+}B^{-}$ and $B^{0}\bar{B}^{0}$ can be taken into account through one
effective $(B\bar{B})_{0}$ threshold whose mixing
potential with $b\bar{b}$ contains an additional factor $\sqrt{2}$ as compared to that for a single
threshold as $B_{s}\bar{B}_{s}$.

\bibliography{botbib}

\end{document}